\documentclass[aps,prl,reprint,nofootinbib,twocolumn,superscriptaddress,showpacs,showkeys,longbibliography]{revtex4-1}
\usepackage{eurosym}
\usepackage{amsmath,amssymb,amstext}
\usepackage[usenames,dvipsnames]{color}
\usepackage{graphicx}
\usepackage{braket}
\usepackage{natbib}
\usepackage{comment}
\usepackage{dcolumn}
\usepackage[english]{babel}
\usepackage{wasysym}
\usepackage[colorlinks,bookmarks=false,citecolor=blue,linkcolor=red,urlcolor=blue]{hyperref}
\begin{document}
\title{   Non-Hermitian topological exciton-polariton corner modes}
\author{Xingran Xu}
\email{thoexxr@hotmail.com}
\affiliation{Division of Physics and Applied Physics, School of Physical and Mathematical Sciences, Nanyang Technological University, Singapore 637371, Singapore}
\author{Ruiqi Bao}
\affiliation{Division of Physics and Applied Physics, School of Physical and Mathematical Sciences, Nanyang Technological University, Singapore 637371, Singapore}

\author{Timothy C.H. Liew}
\email{TimothyLiew@ntu.edu.sg}
\affiliation{Division of Physics and Applied Physics, School of Physical and Mathematical Sciences, Nanyang Technological University, Singapore 637371, Singapore}
\affiliation{MajuLab, International Joint Research Unit UMI 3654, CNRS, Universit\'e C\^ote d'Azur, Sorbonne Universit\'e, National University of Singapore, Nanyang Technological University, Singapore}

\date{\today}

\begin{abstract}
We theoretically study  two-dimensional exciton-polariton lattices and predict that non-Hermitian topological corner modes  can be formed under non-resonant pumping.  As a generalization of the non-Hermitian skin effect, all eigenstates are localized at the two corners in our model. This is also a higher dimensional topology compared to other proposals in exciton-polariton systems and we find that it allows propagating signals in the bulk of the system to travel around defects, which is not possible in one-dimensional topological lattices or two-dimensional lattices with Hermitian edge states. Furthermore, as all polariton states are localized away from an excitation spot, the system offers an opportunity for more accurate measurement of the polariton-polariton interaction strength as the pump-induced exciton-reservoir is spatially separated from {\it all} polariton states.
\end{abstract}
\maketitle

\emph{Introduction.---} Exciton-polaritons are half-light half-matter quasiparticles formed in semiconductor microcavities~\cite{Rev1,Rev2,Rev3,Carusotto2013}. Their matter component offers a significant nonlinearity, while their optical component allows them to be manipulated \cite{Schneider_2016} and observed on optical length scales. Together with their spin degree of freedom ~\cite{Shelykh_2009}, this offers a unique system to study spinor wave behaviour in a controllable and accessible environment. One of the most famous physical phenomena in this system is known as {\it polariton superfluidity}, which is traditionally characterized by the propagation of a polariton wavepacket around a defect without coupling to backscattered states. Shortly after its prediction\cite{Carusotto2004} it was generalized to the spinor case~\cite{Shelykh2006sup} and later observed experimentally \cite{Amo_2009}. The polarization degree of freedom allowed a {\it binary} polariton superfluid behaviour \cite{Cancellieri2012} and modern materials have allowed extension of the behaviour to room temperature \cite{Lerario_2017}.

The absence of coupling to backscattered states is not a unique effect of nonlinearity. Topological Chern systems are also famous for the same claim and well established in topological photonics as well as in polaritonic crystals formed by the etching of semiconductor microcavities into lattice structures  \cite{Klembt_2018}. The difference though in the observable behaviour (aside the underlying physical mechanism) is that the backscattering suppression for a polariton superfluid should occur for a wavepacket propagating in the bulk of the system, while the suppression in a Chern insulator occurs for a wavepacket propagating along the edge of a lattice. The former situation seems more favourable for application of polaritons in coherent exciton-polariton devices~\cite{Fraser_2017}.

While topological Chern systems are an example of Hermitian physics, it should be noted that exciton-polaritons are non-Hermitian systems ~\cite{Gao_2015,Hanai2019}. The application of a non-resonant optical field, which can be spatially patterned or modulated, represents a gain in the system, while the finite lifetime of polaritons represents a natural loss. Non-Hermitian lattices allow non-Hermitian topological effects, such as the skin effect \cite{Helbig_2020,Xiao_2020,Li_2020,Liu2020,Lee2019} where {\it all} the eigenstates of a system become localized at an edge and a new bulk-boundary correspondence requires a generalized Brillouin zone (GBZ) method for its explanation~\cite{Kazuki2019,RobertJan2020,Topoorg,yao2018}. The simplest one dimensional skin effect can be realized in the Hatano-Nelson model  \cite{HNmodel1,HNmodel2}, which assumes a non-reciprocal hopping between lattice sites. The Chern number is replaced with a different topological invariant, namely a winding number describing the angular direction in which eigenenergies encircle a point in the complex plane when the wavevector is scanned across the Brillouin zone. If the skin effect is present, then the eigenenergies calculated for an infinite lattice with periodic beriodic boundary condition (PBC) should also encircle the eigenenergies calculated for a finite lattice with otherwise the same parameters and open boundary condition (OBC) \cite{Li_2020,zhang2021universal,Topoorg}. Recent works have also reported the  $\mathbb{Z}_2$  skin effect \cite{Yoshida2020,yokomizo2020nonbloch,Xu2021} and higher-dimensional skin effects \cite{Luo2019,Liusecond,Okugawa2020,Song_2020,zhang2021universal}. To realize the skin effect, coupled resonant optical waveguides  \cite{Wong_2016,Hafezi_2013,Bahari_2017,Klaas2019} are typically considered for changing the coupling between ring resonators, which can be arranged into lattices \cite{Lin_2021,Zhu2020,Nada2017,huawen2021,Mandal2020}. In exciton-polariton systems, the skin effect has been predicted in a one-dimensional lattice where each site exhibits a polarization splitting \cite{Mandal2020,mandal2021topological} equivalent to a coupling between the two polariton spin components. A circularly polarized optical gain breaks the spin symmetry \cite{Klaas2019} and results in an effective asymmetric coupling between sites.

In this work, we present a scheme for a two-dimensional skin effect in a two-dimensional polariton lattice. Here, all eigenstates are localized at the corners of the system. We define a non-trivial topological invariant, namely a winding number, by identifying an appropriate integration direction in the Brillouin zone. As a first application of the two-dimensional skin effect, we consider the propagation of wavepackets in the bulk of the system and find that they travel around defects, in analogy to polariton superfluids although there is coupling to other forward propagating states.  This does not imply there is any similarity with the physical mechanism of superfluidity; after all, we are considering here linear physics rather than an effect of particle-particle interactions. Nevertheless, as applications of polariton superfluidity \cite{Fraser_2017}  depend on the observable result, namely propagation around a defect, rather than how actually it is achieved, the non-Hermitian skin effect can be considered just as relevant in this context.

As a second application, we note that a critical challenge in the field of exciton-polaritons has been the measurement of the strength of the polariton-polariton interaction strength. This is often inferred from the energy shift of interacting polaritons, however, this is also affected by interactions of polaritons with a reservoir of higher energy states that are also excited. This has resulted in measurements spanning orders of magnitude, even using the same material system \cite{Estrecho2019}. Direct measurements have only been achievable in ultra-high-quality factor microcavities, where a polariton wavepacket lives long enough to be separated from the higher energy states \cite{Estrecho2019}. Using the skin effect, we find that as all eigenstates are localized at the corner of the lattice,  one can readily separate from most of the reservoir,  which in principle allows accurate measurement of the interaction strength even in material systems where ultra-high-quality factor has not been achieved In particular, this includes the growing body of work on polariton lattices and novel materials \cite{Anton_Solanas_2021,Su_2020NP}.

\begin{figure}[h]
\centering
\includegraphics[width=0.48\textwidth]{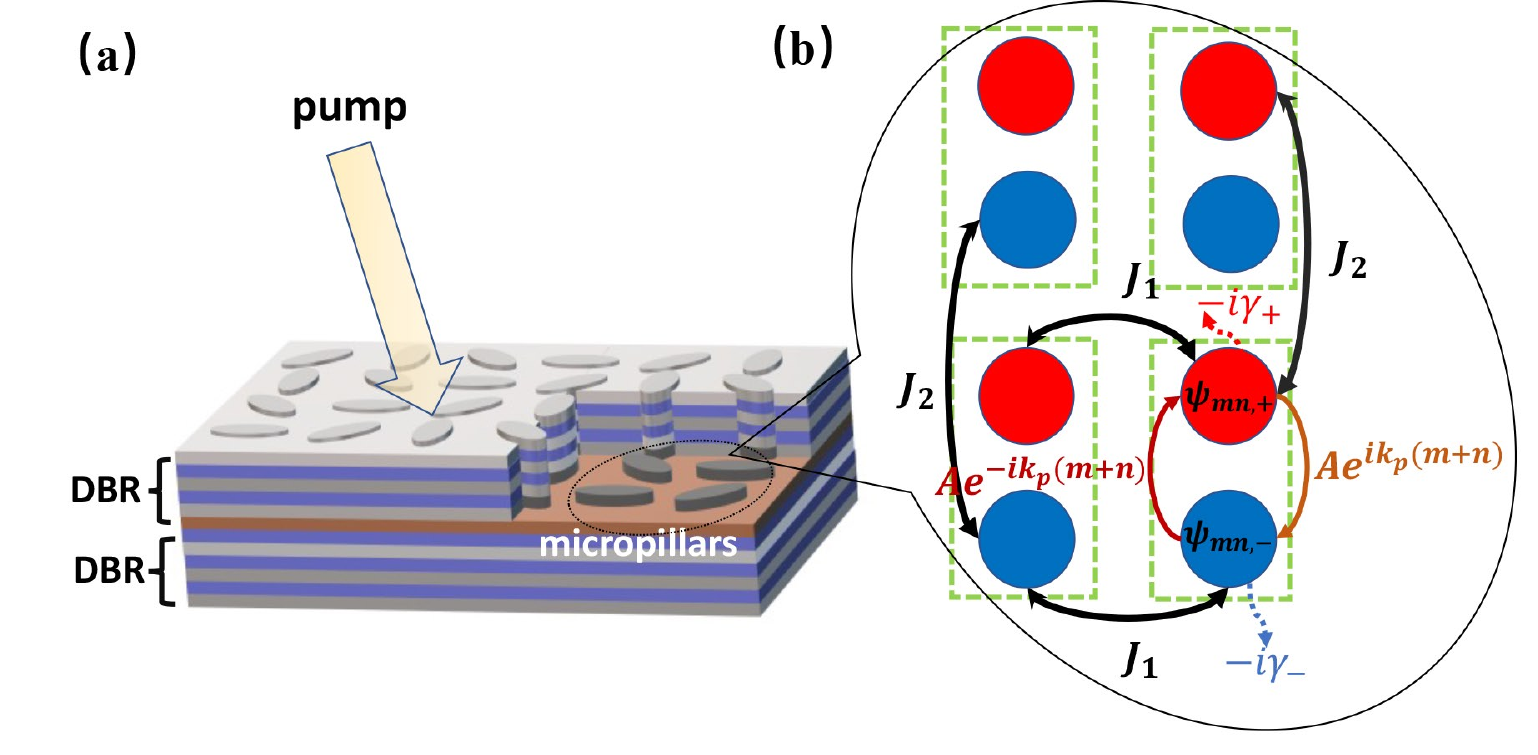}\\
\caption{(a) Schematic representation of a two-dimensional polariton lattice of micropillars with local polarization splitting. (b) Each micropillar supports two different spins of polaritons, illustrated with red and blue circles. The figure shows the different couplings between the spins of neighbouring lattice sites. }\label{scheme}   
\end{figure}

\emph{Model.---} We consider the localization of exciton-polaritons at the sites of a square lattice. As examples, these sites could be formed from the etch and overgrowth technique, etching of complete micropillars, or the milling of mirror layers in an open microcavity design \cite{lackner2021fully}. We further assume that each site exhibits a polarization splitting with a well-defined direction, which has been previously achieved experimentally using elliptical micropillars \cite{Klaas2019,Gerhardt2019}. Within the tight-binding and mean-field approximations, exciton-polaritons are described by a two-component wavefunction driven-dissipative Gross-Pitaevskii equation:
\begin{eqnarray}
i\hbar\frac{\partial\psi_{m,n,\pm}}{\partial t}&=&J_{1}\left(\psi_{m,n-1,\pm}+\psi_{m,n+1,\pm}\right)\nonumber\\
&+&J_{2}\left(\psi_{m+1,n,\pm}+\psi_{m-1,n,\pm}\right)+Ae^{\pm i\theta_{m,n}}\psi_{m,n,\mp}\nonumber\\
&+&\left[\frac{i\hbar}{2}(RN_{m,n,\pm}-\gamma_{C})\right]\psi_{m,n,\pm}\nonumber\\
&+&\left(g\left|\psi_{m,n,\pm}\right|^{2}+g_{R}N_{m,n,\pm}\right)\psi_{m,n,\pm},\label{gp}
\end{eqnarray}
where $\psi_{m,n,\pm}$ is the wavefunction of polaritons at the site with indices $(n,m)$ and spin polarization $\pm$. $J_1=J_2$ is the nearest neighbour hopping in the $x$  and $y$ direction, $A$ is the polarization splitting strength, and $\gamma_{C}$ is the decay rate of the polaritons. The hoppings in the unit cell have the position-dependent phase $\theta_{m,n}=k_p (m+n)$   which is determined by the direction of polarization splitting at each site (that is, the orientation of elliptical micropillars).  $g$ is the nonlinear interaction between polaritons and $g_R$ is the interaction between the polaritons and reservoir excitons. The reservoir excitons can themselves be spin polarized and excited by a non-resonant pump. The reservoir exciton density $N_{m,n,\pm}$ at site $(n,m)$ with spin $\pm$ is described by the rate equation:
\begin{equation}
\frac{\partial N_{m,n,\pm}}{\partial t}=P_{\pm}-\left(\gamma_{R}+R\left|\psi_{m,n,\pm}\right|^{2}\right)N_{m,n,\pm},\label{req}
\end{equation}
where $P_\pm$ is the pump strength of the non-resonant laser, $\gamma_R$ is the decay rate, and $R$ is the stimulated scattering strength,  which allows reservoir excitons to scatter into condensed polaritons. The above two equations describe polariton condensation under nonresonant pumping. 




\emph{The non-Hermitian corner modes---} In the steady-state (represented here with superscript 0), the polariton and reservoir densities are related by:
\begin{equation}
N^0_{m,n,\pm}=P_{m,n,\pm}/(\gamma_R+R|\psi_{m,n,\pm}|^2).
\end{equation}
We will consider first the low density regime (that is, not much above the threshold for condensation), in which case we can approximate $N^0_{m,n,\pm}\approx P_{m,n,\pm}/\gamma_R$. Considering a spatially uniform pumping, the gain term in Eq.  (\ref{gp}) becomes also spatially uniform and we can define constant effective decay rates for the two polarizations as: $\hbar/2(\gamma_C-R N_{m,n,\pm}) = \gamma_\pm$. While these approximations are useful, for simplifying the calculation of the bandstructure, we will return later to the full numerical simulation of the coupled reservoir dynamics. In any case, note that there can be a difference in the effective decay rates of the different spins, which is determined by the polarization of the non-resonant laser. 

\begin{figure}[h]
\centering
\includegraphics[width=0.5\textwidth]{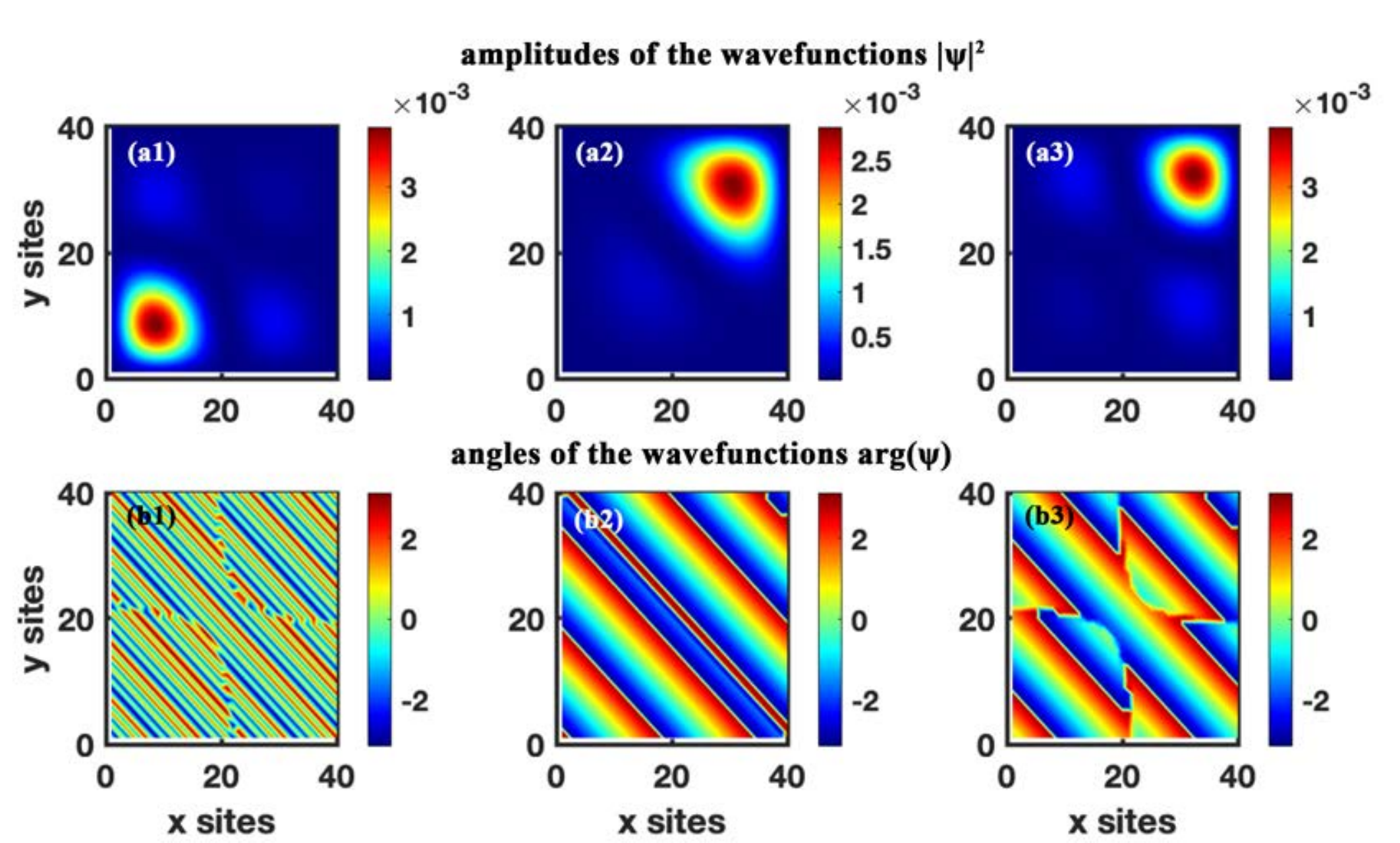}\\
\caption{Amplitudes (the first row)  and the angles (the second row) of three selected eigenfunctions  of Hamiltonian with OBC.  Parameters: $2J_1/\hbar\gamma_C$=$2J_2/\hbar\gamma_C$=$1$, $k_p =\pi/4$,  $\gamma_+/\gamma_C$=$0.1$, $\gamma_-/\gamma_C$=$0.8$, and $2A/\hbar\gamma_C $=$4$.}\label{wf}   
\end{figure}
Furthermore, in the low density regime, we can neglect the last two terms in Eq. (\ref{gp}) and the system can be described with an effective Hamiltonian with only the linear couplings illustrated (see the supplementary material) with only the linear couplings illustrated in Fig. \ref{scheme}(b).  We take $40\times 40$ sites as an example and the OBC is considered in Fig. \ref{wf}. All wavefunctions are localized at the first or the last corners in Figs. \ref{wf}(a1)-(a3),  which we interpret as a higher dimensional non-Hermitian skin effect compared to the corner state in the Hermiitian condition \cite{Rimi2020,Kim_2020,Zhong_2021,Kirsch_2021} .  Meanwhile, the corresponding phases are shown in Figs. \ref{wf}(b1)-(b3),  with a clear gradient along anti-diagonal directions..

An established characteristic of non-Hermitian topology is that the eigenenergies  in the system with in PBCs can form a loop in the complex energy spectrum.  To prove this, we use Fourier transformation to derive the PBC Hamiltonian in reciprocal space as
\begin{equation}
H_{k_x,k_y}=\left(\begin{array}{cc}
 h(k_x+k_p,k_y+k_p)-i\gamma_{+} & A\\
A & h(k_x,k_y)-i\gamma_{-}
\end{array}\right),\label{pbc}
\end{equation}
with $h(k_x,k_y)=2J_{1}\cos\left(k_{x}\right)+2J_{2}\cos\left(k_{y}\right)$  the hopping energy in different directions with wavevectors $k_x$ and $k_y$. The above Hamiltonian is written in the basis of $\pm$ spins and found it convenient to shift the origin of the wavevectors by $k_p$.

\begin{figure}[h]
\centering
\includegraphics[width=0.5\textwidth]{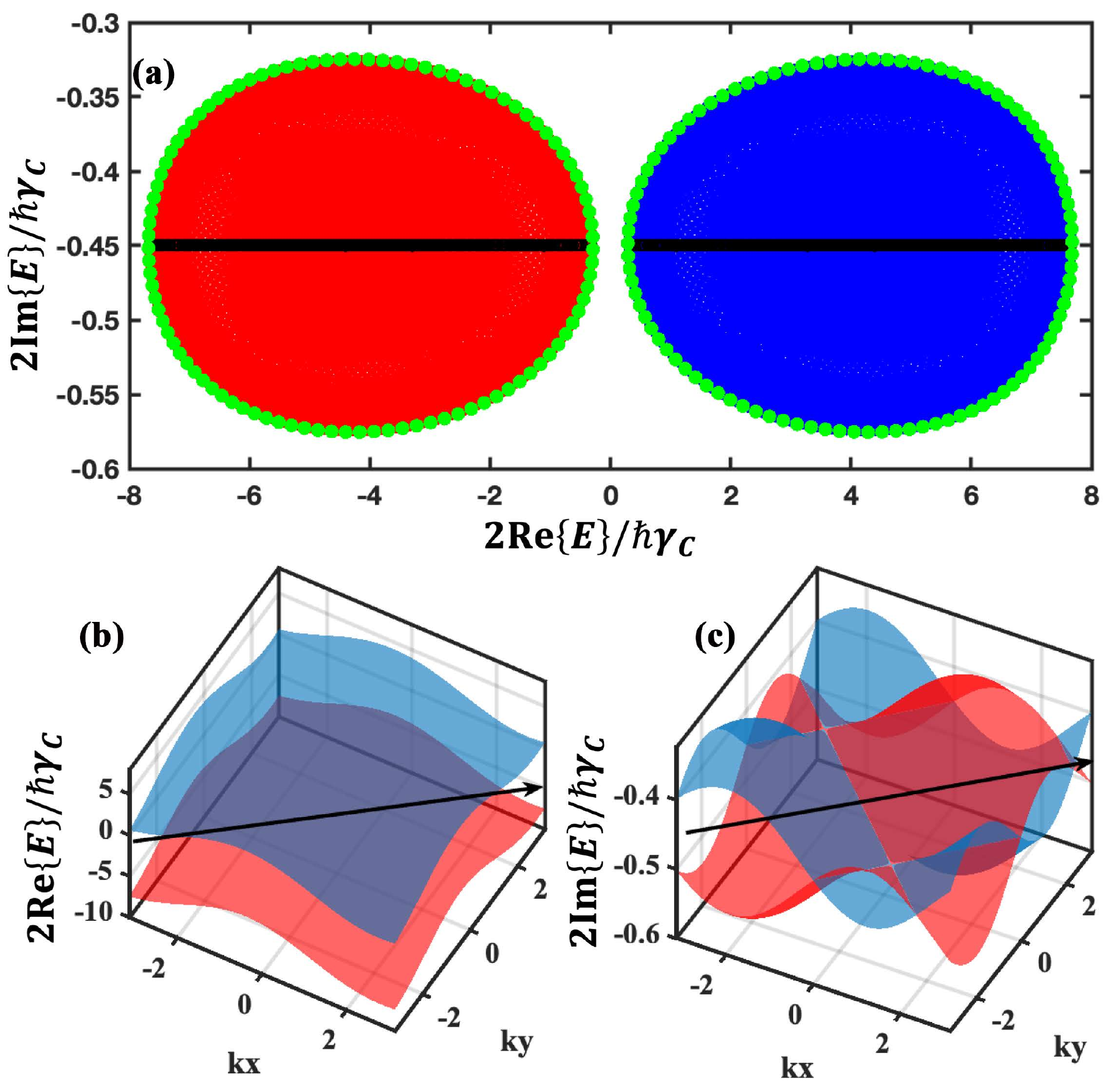}\\
\caption{Complex energy spectrum with PBC and OBC (a),  and real energy (b) and imaginary energy band (c) of Eq. (\ref{pbc}) The black lines are the OBC energies and the blue and red areas are filled with the PBC energies of upper and lower bands. Meanwhile, the green dots are the complex energies along with the selected integration direction for defining the winding number in (a). Parameters: $2J_1/\hbar\gamma_C$=$2J_2/\hbar\gamma_C$=$1$, $k_p =\pi/4$,  $\gamma_+/\gamma_C$=$0.1$, $\gamma_-/\gamma_C$=$0.8$, and $2A/\hbar\gamma_C $=$4$. }\label{ES}   
\end{figure}

As is shown in Fig. \ref{ES} (a), the whole energy spectrum is symmetric about the imaginary axis,  which is corresponding with the two localizations in Figs. \ref{wf}(a1)-(a3). The OBC energy spectrum (black lines) is inside the 2D PBC energy spectrum (blue and red areas) calculated by Eq. (\ref{pbc}). Meanwhile,  the OBC energies are encircled by the complex energy spectrum along with a selected integration direction in $\vec{k}$ (green dots), which will allow us to define a winding number. The imaginary parts of all OBC eigenenergies are $-i(\gamma_++\gamma_-)/2$. The different decay rates break the Hermiticity in a non-trivial way and generate the non-Hermitian corner modes  The winding number of the eigenenergies can be calculated by \cite{Topoorg,Kawabata20202d,Leykam2017}
\begin{eqnarray}
W=\frac{1}{2\pi i}\int_{k'}\frac{d}{d k'}\log\left[Det(H-E_R)\right] d k',
\end{eqnarray}
where $E_R$ is the complex reference energy and we choose $\vec{k}'=(k',k')/\sqrt{2}$ as the integration  direction crossing the whole Brillouin zone. 

The winding number can be defined with a choice of many integration directions, however only the selected direction $\vec{k}$ can encircle all OBC energies and identify the non-trivial topology with non-zero result. The integration direction is also corresponding  with the momentum shift direction in the Hamiltonian.

\emph{The specific direction signal propagation---}
The skin effect can make all wave functions have non-reciprocal flux to a boundary. In the 1D case, the HN model shows enhanced propagation in one direction and weakened propagation in the other.  However, if there is a defect in the lattice, the signal has a very low chance to cross the defect and most of the propagation will stop at the defect (aside a small amount allowed by quantum tunnelling).  The higher dimensional non-Hermitian corner modes give more flexibility for the signal to propagate efficiently and allow direction to be controlled.  Defects can be circumvented and do not stop propagation.

\begin{figure}[h]
\centering
\includegraphics[width=0.5\textwidth]{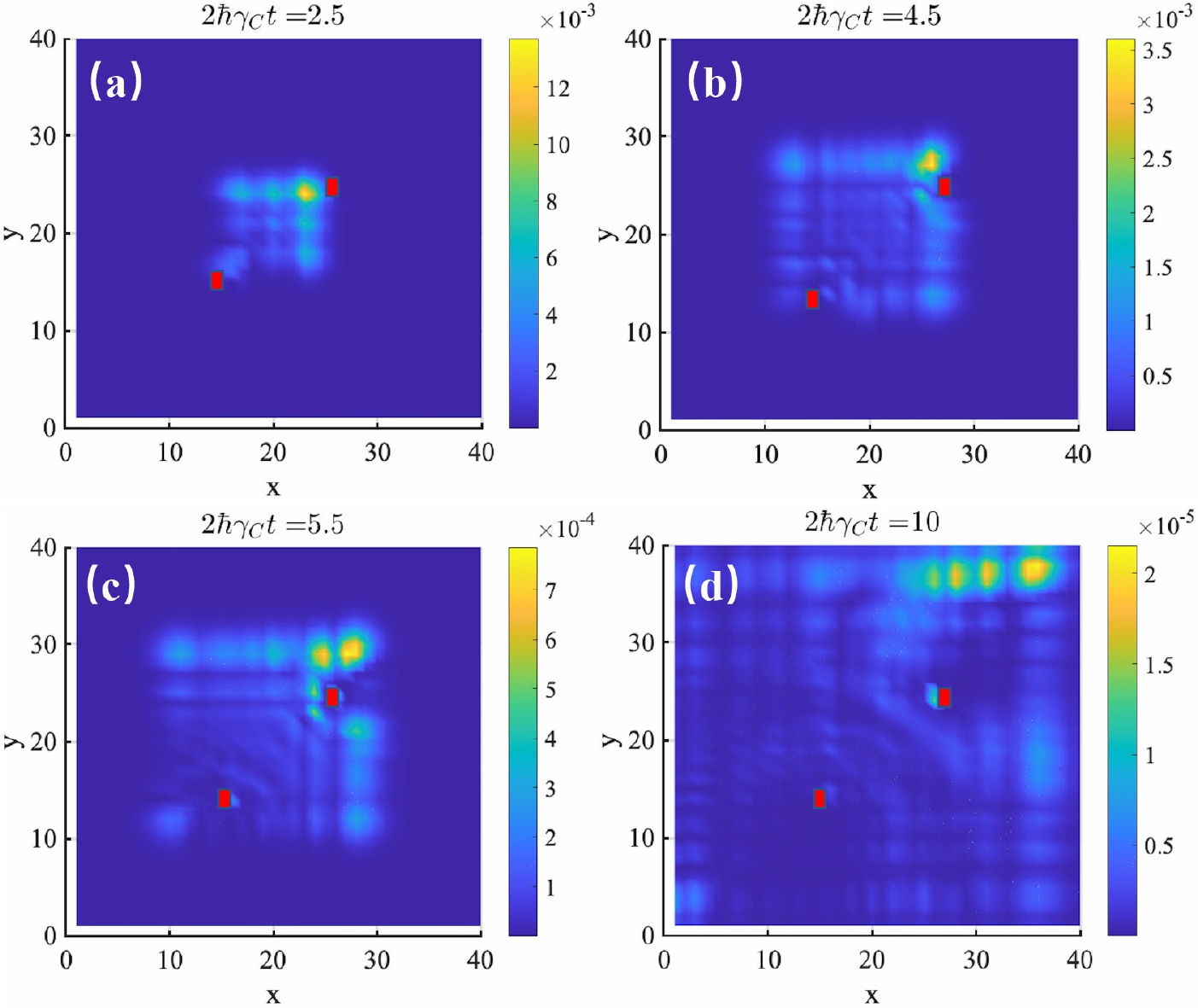}\\
\caption{Time evolution of (in the linear regime) polaritons with two defects at $(15,15)$ and $(25,25)$ sites. Parameters: $2J_1/\hbar\gamma_C$=$2J_2/\hbar\gamma_C$=$1$, $k_p =\pi/4$,  $\gamma_+/\gamma_C$=$0.1$, $\gamma_-/\gamma_C$=$0.8$, and $2A/\hbar\gamma_C $=$4$. }\label{defv}   
\end{figure}

The dynamic behaviors of the polariton condensates are shown in Fig.~\ref{defv}. There are two defects (red dots) in the lattice with potential strength $2V_d/\hbar\gamma_C=10$. The initial states are prepared in the center of the lattice scheme and the linear approximation is considered.  In experiment the initial wavepacket could be injected with a focused laser pulse, while maintaining also the polarized and spatially uniform non-resonant continuous laser excitation needed to achieve $\gamma_+\neq\gamma_-$. The eigenstates of the wavefunctions have two localizations as is shown in Figs. \ref{wf}(a1)-(a2), however, their decay rate is different. When we release the condensates they will go to the two corners automatically. The condensation will decay faster when it goes to the first site $(1,1)$ and decay slower when it goes to the last site $(L,L)$. The defects only have a little influence on the propagation as is illustrated in Fig. \ref{defv}(d) and most of the condensates propagate around the defects.

In addition to allowing propagation around a defect, an advantage of using a 2D lattice is that one has access to direction control. In particular, by varying the phases in the spin coupling term, one can change the direction of favoured propagation. In principle, it would also be possible to design a lattice that directs all polaritons to a specific site.

\emph{The  interaction measurement---}Despite its importance as a key paramter in polariton physics, the strength of polariton-polariton interaction is not easy to distinguish experimentally. One typically hopes to infer it from a nonlinear shift of the polariton energies, however, a severe complication is that the shift in energy of a polariton condensate can be associated to two terms:
\begin{equation}
V_{c}=\frac{\sum_{m,n,\pm}g|\psi_{m,n,\pm}|^{4}}{\sum_{m,n,\pm}\psi_{m,n,\pm}^{*}\psi_{m,n,\pm}},
\end{equation}
and 
\begin{equation}
V_{r}=\frac{\sum_{m,n,\pm}\psi_{m,n,\pm}^{*}g_RN_{m,n,\pm}\psi_{m,n,\pm}}{\sum_{m,n,\pm}\psi_{m,n,\pm}^{*}\psi_{m,n,\pm}}.
\end{equation}
The first represents the interaction strength due to polariton-polariton interaction, while the second represents the interaction strength due to polariton-reservoir interaction,  normalized by the total condensate density. The density of the polaritons and the density of the reservoir have different behaviors with the increasing pump, however, it is challenging to separate the two contributions and experimental reports of interaction strengths have spanned multiple orders of magnitude even when considering the same material systems \cite{Estrecho2019}. So far, the best method of solving this problem has been to use a long lifetime microcavity in which the polariton condensate initially excited at the pump spot position moves to a location far from the pump spot such that the $V_r$ term goes to zero. This requires  a successive coupling of modes with overlap with the pump spot, in which the condensate initially forms, to those delocalized from the pump spot, possibly via energy relaxation processes. While possible in long lifetime microcavities, this is not possible in more common samples with limited lifetime. In our considered system though, {\it all} modes can be delocalized from the pump spot as the skin effect ensures their localization in the corner of a lattice.

\begin{figure}[h]
\centering
\includegraphics[width=0.5\textwidth]{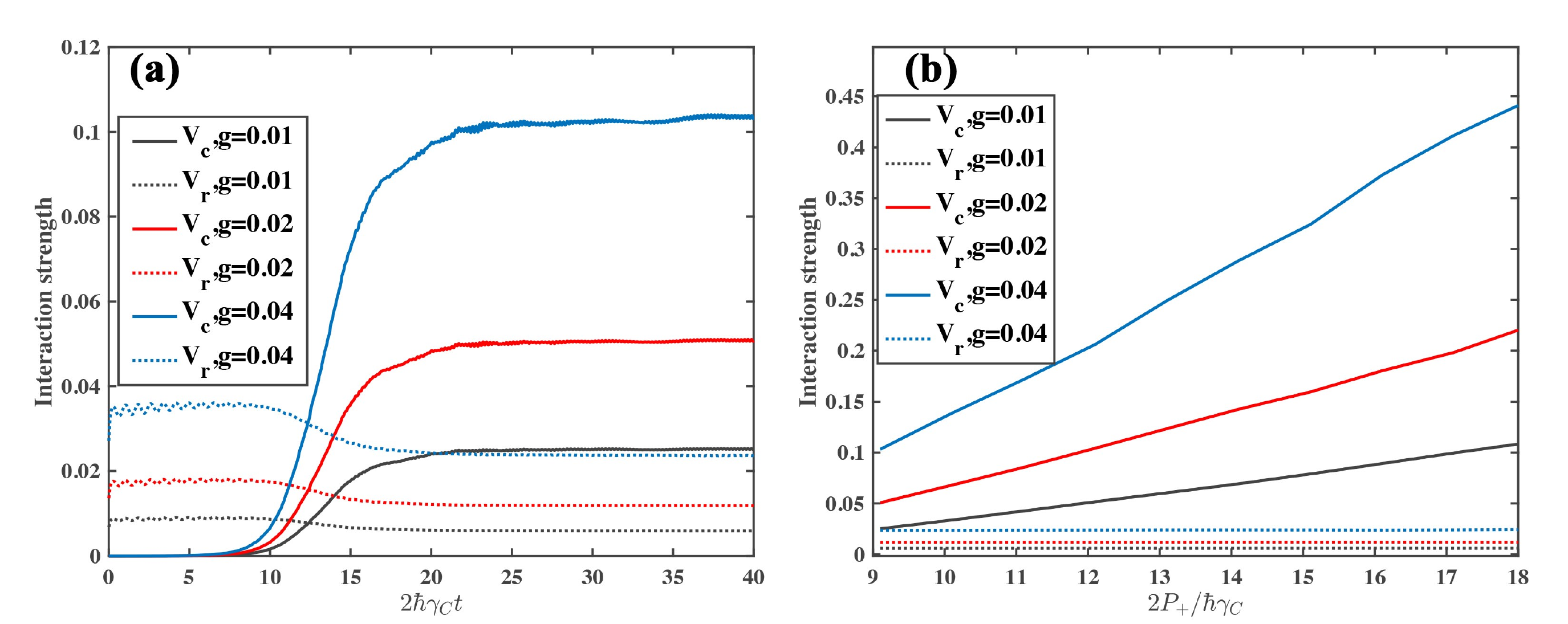}\\
\caption{Time evolution of the polariton-polariton interaction and reservoir-induced interaction (a) and the variation of their steady state values with pumping rate (b).   Parameters : $2J_1/\hbar\gamma_C$=$2J_2/\hbar\gamma_C$=$1$, $k_p =\pi/4$,  $\gamma_R/\gamma_C$=$15$, $R/\gamma_C$=$3$, $2A/\hbar\gamma_C $=$4$, and $P_+=2.5P_-=12.51$ in (a).}\label{Vtp}   
\end{figure}

To compare two types of interaction strength, we keep all terms in Eqs. (\ref{gp})-(\ref{req}) and evolve the coupled equations to find the shape of the stationary state of the condensate.  We consider excitation with a continuous non-resonant pump that allows $\gamma_+\neq\gamma_-$ and evolve an initially random state until a stationary state appears.  As is shown in Fig.~\ref{Vtp}(a), the polariton-polariton interaction will increase from zero and then reach a stationary state.  When the pump rate is fixed, both two interactions will decrease with the increase of the nonlinear strength $g$.  The sensitivities of the two interaction strengths, $V_c$ and $V_r$  to the pump rate are shown in Fig.~\ref{Vtp}(b); the polariton-reservoir interaction is not sensitive to the pump rate and is almost unchanged. In contrast, the polariton-polariton interaction will linearly increase along with the pump rate. Therefore the two forms of interaction can be readily distinguished with power dependent measurements of the polariton energy. Note that this conclusion holds across a reasonable range of parameters $g$ and $g_R$. The reason that $V_r$ does not change with density is because all polariton modes are localized while the reservoir is delocalized.

For not too large density, taking $P_-\propto P_+$, we find empirically that $|\psi_{n,m,\pm}|^2 = c (P_+- P_{th}) |f_{n,m,\pm}|^2$, where $c$ and $P_{th}$ are constants and $f_{n,m,\pm}$ describes the shape of the polariton stationary state, normalized such that $\sum_{m,n,\pm}|f_{m,n,\pm}|^2=1$.  In other words, we assume that for not too large density a change in density does not change significantly the spatial shape of the polariton stationary state. It is not surprising that $|\psi_{n,m,\pm}|^2$ grows linearly with the pump power as this result is known analytically for the case of a spatially uniform condensate, and the coupling between different sites (or spins) does not change the number of particles. $P_{th}$ is the (measureable) threshold pump intensity, at which $|\psi_{n,m,\pm}|^2=0$. Writing the total energy shift as $V_c+V_r$, we have the measureable slope of the energy with power:
\begin{equation}
\zeta = \partial (V_c+V_r)/\partial P_+ \approx \partial V_c/\partial P_+ = g c \sum_{m,n,\pm}|f_{m,n,\pm}|^4.
\end{equation}
The quantity $\sum_{m,n,\pm}|f_{m,n,\pm}|^4$ is readily calculated in theory or can also be measured from the intensity distribution of polaritons in space. To determine g, one then needs to find the constant $c$. In experiment, this could be done with knowledge of the physical value of the polariton density, using the equation $c = \sum_{n,m,\pm} |\psi_{n,m,\pm}|^2/(P_+- P_{th})$. 

Alternatively, note that if we had a periodic boundary condition, equal pumping of all sites, and no coupling of spins, then we would have $|\psi_{n,m,\pm}|^2 = (P_\pm-P_{th})/\gamma_C$.   The presence of coupling between spins and the open boundary condition can make $|\psi_{n,m,\pm}|^2$ spatially non-uniform, however, it would not change the quantity:
\begin{equation}
\sum_{n,m,\pm} |\psi_{n,m,\pm}|^2 = \frac{M(P_+ + P_- - 2P_{th})}{\gamma_C}
\end{equation}
where $M$ is the total number of lattice sites. This is because the coupling between sites and polarizations is itself Hermitian, conserving the total number of particles. Consequently, we can write:
\begin{equation}
c = \frac{M (P_+ + P_- - 2P_{th})}{\gamma_C (P_+ - P_{th})},
\end{equation}
and
\begin{equation}
g = \frac{\zeta \gamma_C (P_+ - P_{th})}{(P_+ + P_- -2P_{th})\sum_{m,n,\pm}|f_{m,n,\pm}|^4},
\end{equation}
where the interaction strength can be obatined by the  fitted coefficients of the interaction-pump curve.

\emph{Discussion---} We have discussed  spinor exciton-polaritons under a polarized nonresonant pump and extended the 1D non-Hermitian skin effect to a two-dimensional effect. The different effective decays of the different spin components of the polaritons and the position-dependent phase hopping in the lattice leads to the localization of all eigenstates in two corners, which we call the non-Hermitian corner modes. By linear approximation, we analyzed the topology of the system and calculated a non-trivial topological invariant, namely the winding number. This requires a judicious choice of integration across the Brillouin zone, which allows covering a set of eigenenergies attained with periodic boundary conditions that fully encircle the eigenenergies attained in the limit of open boundary conditions.

Meanwhile, we give two applications of the higher non-Hermitian topology. One is an effect where propagating polariton wavepackets travel around defects. This may be seen as a somewhat analogous behaviour to that of polariton superfluids, however, we stress that here a propagation direction is defined by the topology of the system and it is an effect of linear rather than nonlinear physics.  Furthermore, we find that as in the skin effect, {\it all} the modes of the system can be localized away from a dominant pumping spot, the system offers an opportunity to more accurately access the polariton interaction strength, even in systems with limited lifetime. The polariton-polariton interaction will linearly increase with the pump rate, while the polariton reservoir interaction  experiences negligible change and so does not get confused with the former. The interaction strength is then readily extracted by the slope of the energy-pump curve.

\emph{Acknowledgements---}We thank Huawen Xu for stimulated discussing. This work was supported by the Singaporean Ministry of Education, via the Tier 2 Academic Research Fund project MOE2018-T2-2-068.

\bibliography{mybib}

\clearpage

\onecolumngrid
\begin{center}
\textbf{\large Supplemental Materials for Non-Hermitian topological exciton-polariton corner modes }
\\
(Dated: \today)
\end{center}

\setcounter{equation}{0}
\setcounter{figure}{0}
\setcounter{table}{0}
\setcounter{page}{1}
\makeatletter
\renewcommand{\theequation}{S\arabic{equation}}
\renewcommand{\thefigure}{S\arabic{figure}}
\renewcommand{\bibnumfmt}[1]{[S#1]}
\renewcommand{\citenumfont}[1]{S#1}
\setcounter{section}{0}
\renewcommand{\thesection}{S\arabic{section}}

\section{S1. The non-Hermitian corner modes}
In this section, we will discuss more the topological properties of the system and give the method to calculate the winding number. As is illustrated in the main text, we use $a_n$ and $b_n$ to represent different spins of the polaritons and the Hamiltonian in real space can be written as
\begin{eqnarray}
H_{m,n}&=&\sum_{m,n}\left[J_{1}\left(a_{m,n}^{\dagger}a_{m,n+1}+b_{m,n}^{\dagger}b_{m,n+1}+h.c.\right)+J_{2}\left(a_{m+1,n}^{\dagger}a_{m,n}+b_{m+1,n}^{\dagger}b_{m,n}+h.c.\right)\right] \nonumber\\&+&\sum_{m,n}\left[-i\gamma_{+}\left(a_{m,n}^{\dagger}a_{m,n}\right)-i\gamma_{-}\left(b_{m,n}^{\dagger}b_{m,n}\right)\right]+\sum_{m,n}\left[Aa_{m,n}^{\dagger}b_{m,n}e^{-ik_{p}\left(m+n\right)}+Ab_{m,n}^{\dagger}a_{m,n}e^{ik_{p}\left(m+n\right)}\right],
\end{eqnarray}
with $\gamma_+$ and $\gamma_-$ are the effective decays if we ignore the dynamics of the reservoir and substitute the stationary density to the GP equation with $\gamma_\pm=\frac{i\hbar}{2}(R N_{\pm}^0-\gamma_{C})$. The periodic boundary energy can be obtained by the Fourier transformation
\begin{eqnarray}
\left\{  \begin{array}{c}
a_{m,n}=\sum_{k_{x}}\sum_{k_{y}}a_{kx,ky}e^{ik_{x}n+ik_{y}m}\\
b_{m,n}=\sum_{k_{x}}\sum_{k_{y}}a_{kx,ky}e^{ik_{x}n+ik_{y}m}
\end{array}\right .
\end{eqnarray}
so
\begin{eqnarray}
H\left(k_{x},k_{y}\right)&=&\sum_{k_{x},k_y}\left[2J_{1}\cos k_{x}\left(a_{k_{x},k_{y}}^{\dagger}a_{k_{x},k_{y}}+b_{k_{x},k_{y}}^{\dagger}b_{k_{x},k_{y}}\right)+2J_{2}\cos k_{y}\left(a_{k_{x},k_{y}}^{\dagger}a_{k_{x},k_{y}}+b_{k_{x},k_{y}}^{\dagger}b_{k_{x},k_{y}}\right)\right]\nonumber\\
&+&\sum_{k_{x},k_y}\left[-i\gamma_{+}a_{k_{x},k_{y}}^{\dagger}a_{k_{x},k_{y}}-i\gamma_{-}b_{k_{x},k_{y}}^{\dagger}b_{k_{x},k_{y}}\right]+A\sum_{k_{x},k_y}\left[a_{k_{x},k_{y}}^{\dagger}b_{k_{x}-k_{p},k_{y}-k_{p}}+a_{k_{x}+k_{p},k_{y}+k_{p}}b_{k_{x},k_{y}}^{\dagger}\right].
\end{eqnarray}
We can shift the momentum of operator $a_{k_{x},k_{y}}^{\dagger}b_{k_{x}-k_{p},k_{y}-k_{p}}$ to $a_{k_{x}+k_{p},k_{y}+k_{p}}^{\dagger}b_{k_{x},k_{y}}$ for the periodic boundary condition. Therefore, the system can be written into periodic Hamiltonian $V^{\dagger}H_{k}V$ with operator vector $V=\left(a_{k_{x}+k_{p},k_{y}+k_{p}},b_{k_{x},k_{y}}\right)^{\top}$ and 
\begin{eqnarray}
H_{k_x,k_y}=\left(\begin{array}{cc}
2J_{1}\cos\left(k_{x}+k_{p}\right)+2J_{2}\cos\left(k_{y}+k_{p}\right)-i\gamma_{+} & A\\
A & 2J_{1}\cos\left(k_{x}\right)+2J_{2}\cos\left(k_{y}\right)-i\gamma_{-}
\end{array}\right).
\end{eqnarray}
The eigenvalues of the system are
\begin{eqnarray}
E_{1,2}&=&-\frac{i}{2}\left(\gamma_{1}+\gamma_{2}\right)+2J_{1}\cos k_{x}'\cos\frac{k_{p}}{2}+2J_{2}\cos k_{y}'\cos\frac{k_{p}}{2}\nonumber\\
&\pm&\sqrt{A^{2}-\delta^2+4\sin\left(\frac{k_{p}}{2}\right)\left[i\delta+J_{1}\sin k_{x}'\sin\left(\frac{k_{p}}{2}\right)+J_{2}\sin k_{y}'\sin\left(\frac{k_{p}}{2}\right)\right]\left[J_{1}\sin k_{x}'+J_{2}\sin k_{y}'\right]}
\end{eqnarray}
with $\delta=(\gamma_+-\gamma_-)/2$ and $k_{x(y)}'=k_{x(y)}+k_p/2$.

The polarization splitting strength $A$ allows coupling of states of different momentum and decay and can open an imaginary gap to form a loop the complex energy spectrum. If  the decay difference $\delta=0$, the eigenenergies will be
\begin{eqnarray}
E_{1,2}=-i\gamma+2J_{1}\cos k_{x}'\cos\frac{k_{p}}{2}+2J_{2}\cos k_{y}'\cos\frac{k_{p}}{2}\pm\sqrt{A^{2}+4\sin\left(\frac{k_{p}}{2}\right)^{2}\left[J_{1}\sin k_{x}'+J_{2}\sin k_{y}'\right]^{2}},
\end{eqnarray}
and $A^{2}+4\sin\left(\frac{k_{p}}{2}\right)^{2}\left[J_{1}\sin k_{x}'+J_{2}\sin k_{y}'\right]^{2}\geq0$. Therefore, the PBC energy spectrum  can not form a loop and  there are no skin modes.

Analogous to the 1D case, the winding number for the eigenenergies is well defined by  
\begin{eqnarray}
W=\frac{1}{2\pi i}\int_{k'}\frac{d}{dk'}\log\left[Det(H-E_R)\right] dk',
\end{eqnarray}
with $k'$  the selected integrated direction. As is illustrated in Fig. 3(a) of the main text, the green color area is the 2D PBC energy spectrum, and only the boundary of this area (green dots) is useful to calculate the winding number. Otherwise, some OBC energies (black lines) can not be encircled if we choose the wrong direction. Hopefully, the origin of the non-Hermitian topology comes from the position-dependent hopping with polarization phase  and decay.


\section{S2. The evolution of 1D polaritons with defects}
In this section, we will compare the evolution of the polaritons in 1D and 2D and discuss the advantages of the higher dimensional topology.

\begin{figure}[h]
\centering
\includegraphics[width=0.9\textwidth]{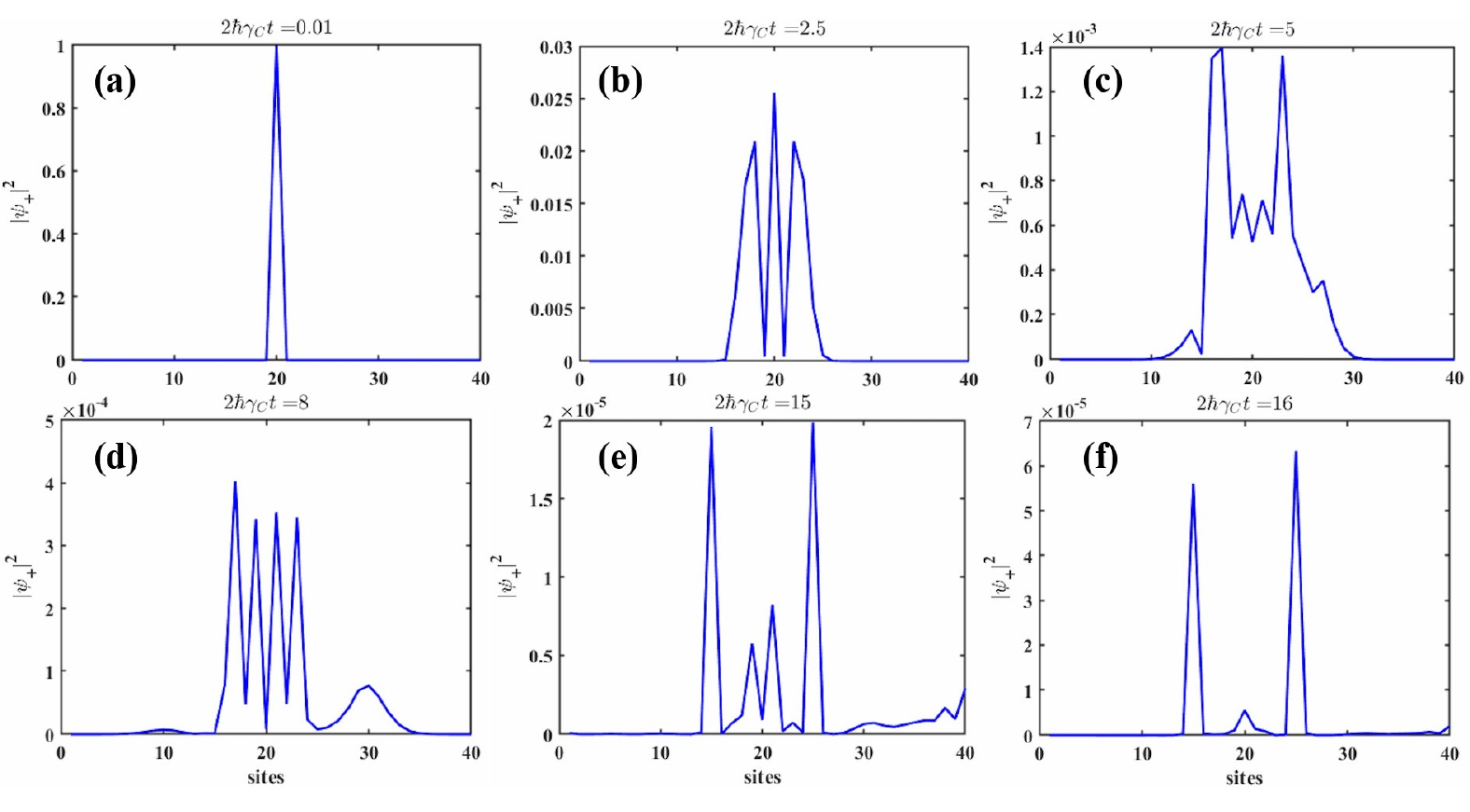}\\
\caption{Time evolution of the 1D polaritons with the strength of the two defects is $2V_d/\hbar/\gamma_C$=10 at $15$ and $25$ sites.  Parameters : $2J_1/\hbar\gamma_C$=$1$, $k_p =\pi/4$,  $\gamma_1/\gamma_C$=$0.1$, $\gamma_2/\gamma_C$=$0.8$, and $2A/\hbar\gamma_C $=$4$.}\label{1D}   
\end{figure}
As is shown in Fig. \ref{1D}, we set $J_2$ to zero to study how the 1D polaritons evolve in the presence of the defects. The initial state is prepared in the center of the lattice as is shown in Fig. \ref{1D}(a) and then the density will go to the two defects (at 15 and 25 sites). The defects will stop most of the propagation, however, the quantum tunneling can let some polaritons across the defects as is shown in Fig. \ref{1D}(d). The polaritons have more chance to cross the right defect than the left one. The decays of the different directions are different. Therefore, the polaritons that go to the left vanish fast while the polaritons can go to the right boundary as is shown in Figs. \ref{1D}(d)-(e). Finally, all polaritons that cross the defects vanish and exist at the defect as is shown in Fig.\ref{1D}(f).

The evolution of polaritons in the 2D lattice is illustrated in Fig. 4 in the main text. The higher dimension gives the polaritons a new path to cross the defects and almost without the density loss. On the other hand, the direction of the propagation can be designed in two dimensions if we can set the position-dependent phase of the hopping (which is defined by the orientations of elliptical micropillars in our proposed scheme).  For example, we can cut off the phase changing in the $y$ direction, and then all wavefunction will localize at the edge of the $x$ boundaryies  instead of  the corners. Another example we can make the phase hoppings (due to the polarization splitting ) in the $x$ direction and $y$ direction are different, so the localized site will also change. So the localized site can be designed to transform the signal to the position we want. Remarkably, even we use  if different phase changing in a different direction, the integrated direction $(1/\sqrt{2},1/\sqrt{2})$ to calculate the winding number will not change because the phase still changes in this direction.

Above all, there are lots of advantages that both 1D and 2D non-Hermitian have like high sensitivity to the signal because of the exceptional points in the energy spectrum. So the 2D corner modes in polaritons can be an ideal detector and information transform tool.

\section{S3. The polariton-polariton interaction measurement}

In this section, we will discuss the polariton-polariton measurement with 2D corner modes in detail. As we mentioned in the main text the equations 2D polaritons with polarization splitting under the nonresonant pump coupled with the reservoir are:
\begin{eqnarray}
i\hbar\frac{\partial\psi_{m,n,\pm}}{\partial t}&=&J_{1}\left(\psi_{m,n-1,\pm}+\psi_{m,n+1,\pm}\right)+J_{2}\left(\psi_{m+1,n,\pm}+\psi_{m-1,n,\pm}\right)+Ae^{\pm i\theta_{m,n}}\psi_{m,n,\mp} \nonumber\\
&+&\left[\frac{i\hbar}{2}(RN_{m,n,\pm}-\gamma_{C})\right]\psi_{m,n,\pm}+\left(g\left|\psi_{m,n,\pm}\right|^{2}+g_{R}N_{m,n,\pm}\right)\psi_{m,n,\pm},\label{gp1}\\
\frac{\partial N_{m,n,\pm}}{\partial t}&=&P_{\pm}-\left(\gamma_{R}+R\left|\psi_{m,n,\pm}\right|^{2}\right)N_{m,n,\pm},\label{req1}
\end{eqnarray}
where, $J_1$ is the nearest hopping in the $x$ direction, $J_2$ is the nearest hopping in the $y$ direction, $A$ is the TE-TM splitting strength, and $\gamma_{C}$ is the decay rate of the polaritons. The hoppings in the unit cell have the position-dependent phase $\theta_{m,n}=k_p (m+n)$ are determined by the polarized direction of TE-TM modes. $g$ is the nonlinear interaction of the condensation and $g_R$ is the interaction between the polaritons and reservoirs. $P_\pm$ is the pumping rate of different spins and $\gamma_R$ is the decay of the reservoir.

\begin{figure}[h]
\centering
\includegraphics[width=0.95\textwidth]{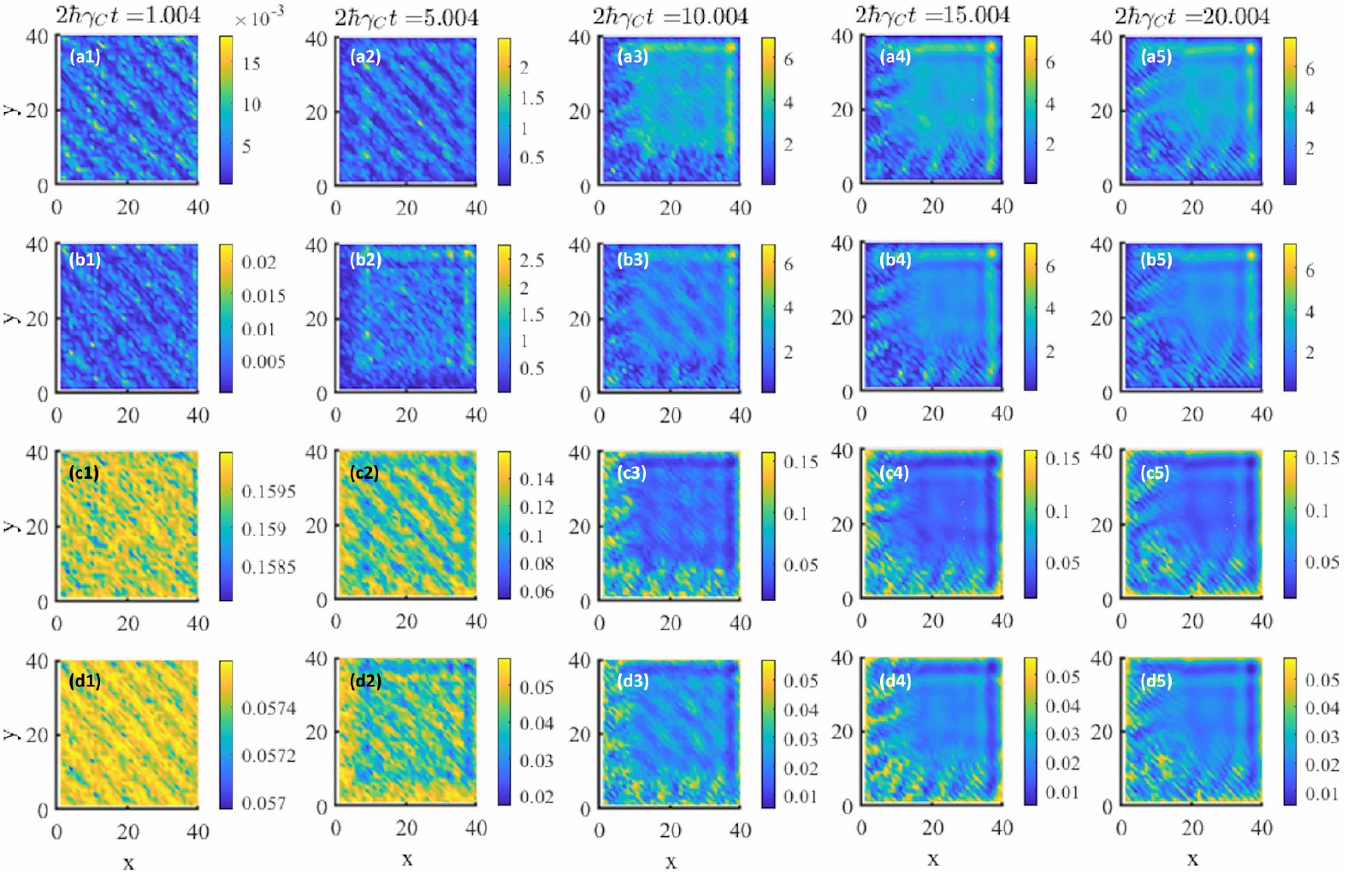}\\
\caption{Time evolution of the density of the spin-up (the first row) and spin-down (the second row) polaritons and the time evolution of the  density of the spin-up reservoir (the third row) and spin-down reservoir (the forth raw). Parameters : $2J_1/\hbar\gamma_C$=$2J_2/\hbar\gamma_C$=$1$, $k_p =\pi/4$,  $\gamma_R/\gamma_C$=$15$, $R/\gamma_C$=$3$, $2A/\hbar\gamma_C $=$4$, $g_R=2g$=$0.04$, and $P_+=2.5P_-=12.51$.}\label{PRT}   
\end{figure}

To measure the polariton-polariton interaction, we use a random function  as the initial state as is shown in the first column in Fig.\ref{PRT} and then evolve Eqs. (\ref{gp1})-(\ref{req1}).  The condensation will go to the localized point automatically as is shown in Figs. ~\ref{PRT}(a3)-(a5) and (b3)-(b5). Meanwhile, the density of the reservoir sinks at that site as is shown in Figs. ~\ref{PRT}(c3)-(c5) and (d3)-(d5).  More than that the densities of the two reservoirs are different which leads to different effective decays of the different spin of the condensation as we mentioned in the linear approximation. In the weak nonlinear region, the topology still works and drives all polariton states to be localized.

\begin{figure}[h!]
\centering
\includegraphics[width=0.8\textwidth]{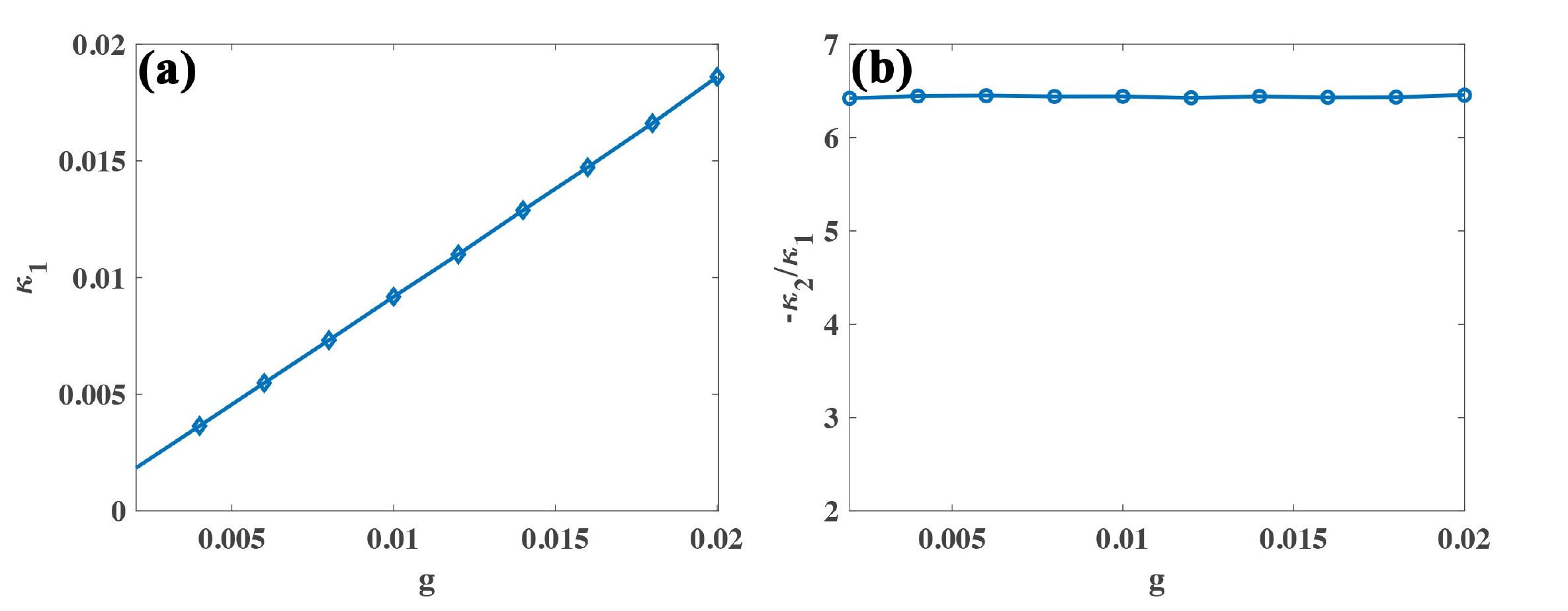}\\
\caption{ Fitted coefficients of the polariton-polariton interaction.  Parameters : $2J_1/\hbar\gamma_C$=$2J_2/\hbar\gamma_C$=$1$, $k_p =\pi/4$,  $\gamma_R/\gamma_C$=$15$, $R/\gamma_C$=$3$,  $g_R=2g$, and  $2A/\hbar\gamma_C $=$4$.}\label{scaling}   
\end{figure}

We calculate two types of interaction: polariton-polariton interaction $V_c$ and the reservoir-polariton interaction $V_r$. $V_r$ is not sensitive to the pump strength and keeps the same value when we increase the pump. However, $V_c$ is almost linearly increased along with the increase of the pump as is shown in Fig. 5(b). We can build the relation $V_c=\kappa_1 P_++\kappa2$ and these two fitted coefficients are shown in Figs. \ref{scaling} (a)-(b). The slope ($\kappa_1$) reveals the interaction strength and is linearly increased with the increase of the polariton-polariton interaction. The intercepts with the pump axis ($-\kappa_2/\kappa_1$) with different interaction strengths are shown in Fig. \ref{scaling} (b) which are unchanged. These intercepts are the threshold of the pump. In Eqs.(\ref{gp1})-(\ref{req1}), the threshold of the both pump is $P_{th}^0=\gamma_C\gamma_R/R=5$, however, we take $P_-=0.4P_+$ so $P_{+,th}$ is larger than $P_{th}^0$. Remarkably,  although $P_{-,th}=0.4P_{+,th}$ is smaller than $P_{th}^0$, this doesn't mean the spin-down polaritons can be prepared under $P_{th}^0$ just because we set the relation between these two pumps. Furthermore, the spin-up and spin-down polaritons have the  polarization splitting coupling. Therefore the spin-down polaritons can be transformed from the spin-up one even under the threshold but the pumping of spin-up polaritons must be larger than $P_{th}^0$.

\end{document}